\documentstyle[titlepage,12pt]{article}
\begin{document}
\def\half{\frac{1}{2}}
\def\sgn{\mbox{sgn}}
\def\be{\begin{equation}}
\def\grad{\nabla}
\def\ee{\end{equation}}
\def\rq#1{(\ref{eq#1})}
\def\rsq#1{(\ref{seq#1})}
\def\to{\rightarrow}
\def\gt{\tilde{g}}
\def\Rt{\tilde{R}}

\newcount\sectionnumber
\sectionnumber=0
\def\be{\begin{equation}}
\def\ee{\end{equation}}

\title{Classical Models of Subatomic Particles}

\author{R.B. Mann and M.S. Morris\\
        Department of Physics\\
        University of Waterloo\\
        Waterloo, Ontario\\
        N2L 3G1}
\date{July 7, 1993\\
WATPHYS TH-93/02}

\maketitle

\begin{abstract}
We look at the program of modelling a subatomic particle---one having mass,
charge, and angular momentum---as an interior solution joined to a classical
general-relativistic Kerr-Newman exterior spacetime. We find that the
assumption of stationarity upon which the validity of the Kerr-Newman
exterior solution depends is in fact violated quantum mechanically for
all known subatomic particles. We conclude that the appropriate stationary
spacetime matched to any known subatomic particle is flat space.
\end{abstract}

The most basic properties of a subatomic particle are its mass $M$,  charge
$Q$ and spin $J$ (and lifetime $\tau$ if it is unstable). As such it would
seem  natural from a general-relativistic  viewpoint to describe the
spacetime metric associated with a given subatomic particle by a
Kerr-Newman metric
\cite{Kerr}
\begin{eqnarray}
ds^2 &=& -dt^2 + \Sigma\left(\frac{dr^2}{\Delta}+d\theta^2\right) +
(r^2+a^2)\sin^2(\theta) d\phi^2 \nonumber \\
&& +  \frac{r {\cal M}}{\Sigma}(dt-a
\sin^2(\theta) d\phi)^2   \label{1}
\end{eqnarray}
at least at distances large compared to the characteristic size $R$ of the
particle. In (\ref{1})
\be
{\cal M} \equiv 2M-Q^2/r \quad \Sigma = r^2 + a^2 cos^2(\theta) \quad
\Delta \equiv r^2 + a^2 - {\cal M}r  \label{2}
\ee
where $a \equiv J/M$ and units are such that $G=c=1$.

For subatomic particles let $J = N_s/2$ and $Q=N_e e$ where $e$ is the
charge of an electron. Assuming that (\ref{1}) provides a valid description
for $r > R$  of the spacetime due a given subatomic particle,  (so that the
effects of strong and weak interactions are neglected) one finds $a/M >>
Q/M >> 1$ for all known quarks, leptons, baryons, mesons and nuclei
(except for spin-zero particles, in which case only the latter inequality
is satisfied). The
first inequality is violated only when  $N_s < 2 N_e\sqrt{\alpha}
\frac{M}{M_{Pl}}$, (where $\alpha$ is the fine-structure constant) whereas
the second is violated when $N_e < \sqrt{\alpha}\frac{M}{M_{Pl}}$. These
situations only hold macroscopically, when the mass of the body is
appreciably larger than the Planck mass $M_{Pl}$, although the first
inequality could be violated for a body with a small mass but a large
charge. Hence for all subatomic particles the metric (\ref{1}), if assumed
valid for all values of $r$, describes the field of a naked singularity.

Unless one is willing to live with such an unattractive scenario
(along with whatever empirical difficulties it may cause),
it is clear that the
Kerr-Newman metric cannot be valid for all values of $r$ for a subatomic
particle; rather it
must be matched on to some interior solution to the
Einstein equations.  Attempts to find such interior solutions have been
carried out ever since the Kerr metric was discovered \cite{Des}.
Such models have been plagued by a variety of unphysical features, including
superluminal velocities and negative mass distributions
\cite{Israel,Hamity,Lopez}.
Indeed some authors \cite{Bonnetal} have advocated that current
data limiting the size of the electron to be smaller than $10^{-16}$cm
(and the assumption that general relativity is valid at these distance
scales) imply that the electron must have a negative rest mass density.

While we regard the search for an interior solution to match onto the
Kerr-Newman metric as being of interest in its own right, we argue here
that assertions concerning the negativity of the rest mass density of the
electron (and all other known elementary particles) are unwarranted.  The
unphysical features described above arise because the matching is  carried
out either at $r=0$ or at half the classical electromagnetic  radius  $r_Q
= Q^2/(2M)$ \cite{Gron}; this effectively results in  ascribing the
electromagnetic rest energy of the particle to be too large relative to
the laboratory value of its rest mass, and so a negative mass density must
be introduced to compensate.   However, the matching conditions must not be
applied at a distance scale below which the Kerr-Newman metric can no
longer be  trusted.  This distance scale can be set by other physics in one
of two ways: either new structure occurs due to the physics of the particle
considered, or the assumption of stationarity breaks down.

The former case occurs for
(stable) baryons and nuclei: for all nuclei (including Hydrogen), one can
easily check that the nuclear radius $r_N\equiv 1.07 A^{1/3}$ fm \cite{Enge}
is larger than either the electromagnetic charge radius or the Kerr parameter
$a$. Matching the exterior Kerr-Newman metric for baryons and nuclei to an
interior solution (whose stress-energy tensor must be determined by nuclear
effects) must therefore take place at a distance $r_M \ge r_N >> a, r_Q$.

For charged leptons, quarks or mesons, the Kerr-Newman solution will not be
applicable if the particle is not stationary. By the uncertainty
principle, the particle would have to be moving at relativistic speed on
distance scales shorter than the Compton wavelength $\lambda_c$ of the
particle, the length scale at which the  average quantum zero-point
kinetic energy of the particle is comparable to its rest energy.  Requiring
that $|\vec{p}| << m$ implies that the matching must take place at $r_M
>> \lambda_c$  ({\it e.g.} $r_M \sim 100 \lambda_c$). This scale is
much larger than the parameter $a = N_s\lambda_c/2$ for all known
non-baryonic subatomic particles.

If the solution were stationary, one could  employ the matching conditions
using the charged Lense-Thirring metric
\begin{eqnarray}
ds^2 &=& -(1-\frac{2M}{r}+
\frac{Q^2}{r^2})dt^2 +  (1+\frac{2M}{r}-\frac{Q^2}{r^2})(dx^2+dy^2+dz^2)
\nonumber \\
&&\quad +
2 \frac{Q^2-2Mr}{r^4}(\vec{x}\times\frac{\vec{J}}{M}) \cdot d\vec{x} dt
\label{2a}
\end{eqnarray}
which is a post-Newtonian solution to the Einstein equations for the metric
exterior to a charged spinning sphere of constant density  where $M/r <<
1$, $J/r^2 << 1$ and $Q^2/r^2 << 1$. Under these conditions, this metric is
equivalent to (\ref{1}) provided $J/M = a$. In principle one could
determine the values of $M$, $J$ and $Q$ for a body by Gaussian integration
of the gravitational field at a large distance $r_M$.

However in order to perform such integrations it is necessary that the body
be confined to a region $r<r_M$. Quantum mechanically the uncertainty
principle requires that such confinement impart a root-mean square momentum
$\Delta p \ge \hbar/r_M$ to the particle, necessitating
corrections to the metric
(\ref{2a}). For a given imparted momentum $\vec{P}$
these corrections modify (\ref{2a}) to be
\begin{eqnarray}
ds^2 &=& -(1-\frac{2M}{r}+
\frac{Q^2}{r^2} - \frac{\vec{x}\cdot\vec{D}}{r^3})dt^2 \nonumber \\
&&\quad +  (1+\frac{2M}{r}-\frac{Q^2}{r^2} +
\frac{\vec{x}\cdot\vec{D}}{r^3})(dx^2+dy^2+dz^2) \nonumber \\
&& \qquad +
(2 \frac{Q^2-2Mr}{r^4}(\vec{x}\times\frac{\vec{J}}{M})
- 4\frac{\vec{P}}{r}) \cdot d\vec{x} dt  \label{3}
\end{eqnarray}
where $r = |\vec{x}|$, where $\vec{D}$ is the gravitational dipole moment
resulting from the particle no longer being at the origin.

For macroscopic bodies such corrections are negligible, but for subatomic
particles this is not the case. The term $\frac{\vec{P}}{r} \sim
\frac{\hbar}{r^2_M}$ which is the same order of magnitude as the
$\frac{2Mr}{r^4}(\vec{x}\times\frac{\vec{J}}{M})$ term in (\ref{3}) since
$J\sim\hbar$.
Similarly $\frac{\vec{x}\cdot\vec{D}}{r^3} \sim M/r_M$ and so it is of the
same order of magnitude as the first term in $g_{00}$ in (\ref{3}).
The charge terms are of order $ Q^2 / { r_M }^2 \sim M r_Q / { r_M }^2
<< M / r_M $, and so, even though there will be corrections to these
terms due to the uncertainty principle introducing  electric and magnetic
dipole moments, the charge terms are already negligible relative to the
mass and spin terms we have kept. Since the root-mean-square quantum
corrections are always of the same magnitude as the largest terms we have
kept in the post-Newtonian expansion, we cannot trust keeping those
classical terms.

Thus we suggest that appropriate matching to the Kerr-Newman geometry
for the electron is constrained by stationarity to take place at
radial distances from the
particle much larger than the Compton wavelength. The interior solution will
be modelled by a quantum distribution. But, however large the matching
radius ($ r_M $) is taken to be, the act of measuring the
spacetime curvature on a
surface at that distance (i.e. measuring the parameters of the Kerr-Newman
metric), would, again by the uncertainty principle,
kick the momentum of the electron by $ \Delta p \ge m \lambda_c / r_M $,
introducing quantum non-stationarity corrections to the metric of
order $ \lambda_c / r_M $. These corrections for an electron are the
same order as
the $ a / r $ term kept even in the Lens-Thirring approximation to
the Kerr-Newman geometry. This means the uncertainty principle
should make it impossible to measure the Kerr-Newman or even the
charged Lens-Thirring parameters, and the appropriate stationary
solution matching a quantum electron is flat.

Unfortunately, this conclusion tends to undermine one good
motivation for trying to model subatomic particles as Kerr-Newman
sources in the first place. This is the fact that the Kerr-Newman metric
predicts the Dirac value of the electron's gyromagnetic ratio. The above
argument, though, would lead us to regard this as a coincidence.

\section*{Acknowledgments}
This work was supported by the Natural Sciences and Engineering Research
Council of Canada.

\end{document}